\documentclass[conference]{IEEEtran}

\usepackage{cite}
\usepackage{color}
\usepackage{multirow}
\usepackage{booktabs,multirow,array}
\usepackage[table]{xcolor} 
\usepackage{ltxcmds}
\usepackage{footnote}
\usepackage{amsthm}

\usepackage[table]{xcolor}
\usepackage{subfigure}
\usepackage{lipsum}
\usepackage{amssymb} 
\usepackage{pdflscape}
\usepackage{afterpage}

\usepackage{pifont}
\newcommand{\xmark}{\ding{55}}%

\ifCLASSINFOpdf
\usepackage[pdftex]{graphicx}
\else
  \usepackage[dvips]{graphicx}
\fi

\begin{document}
\title{A Blockchain-Enabled Incentivised Framework for Cyber Threat Intelligence Sharing in ICS}

\author{\IEEEauthorblockN{Kathy Nguyen$^{\ast}$, Shantanu Pal$^{\ast}$, Zahra Jadidi$^{\ast\mathsection}$, Ali Dorri$^{\ast}$, Raja Jurdak$^{\ast}$}
\IEEEauthorblockA{$^{\ast} $School of Computer Science, Queensland University of Technology, Brisbane, QLD 4000, Australia\\ $^{\mathsection} $School of Information and Communication Technology, Griffith University, Gold Coast Campus, QLD 4222, Australia\\
{thihongthukathy.nguyen@connect.qut.edu.au, (shantanu.pal, zahra.jadidi, ali.dorri, r.jurdak)@qut.edu.au} {}}}




\maketitle


\begin{abstract}

In recent years Industrial Control Systems (ICS) have been targeted increasingly by sophisticated cyberattacks. Improving ICS security has drawn significant attention in the literature that emphasises the importance of Cyber Threat Intelligence (CTI) sharing in accelerating detection, mitigation, and prevention of cyberattacks. However, organisations are reluctant to exchange CTI due to fear of exposure, reputational damage, and lack of incentives. Furthermore, there has been limited discussion about the factors influencing participation in sharing CTI about ICS. The existing CTI-sharing platforms rely on centralised trusted architectures that suffer from a single point of failure and risk companies' privacy as the central node maintains CTI details. In this paper, we address the needs of organisations involved in the management and protection of ICS and present a novel framework that facilitates secure, private, and incentivised exchange of CTI related to ICS using blockchain. We propose a new blockchain-enabled framework that facilitates the secure dissemination of CTI data among multiple stakeholders in ICS. We provide the framework design, technical development and evaluate the framework's feasibility in a real-world application environment using practical use-case scenarios. 
Our proposed design shows a more practical and efficient framework for a CTI sharing network for ICS, including the bestowal and acknowledgment of data privacy, trust barriers, and security issues ingrained in this domain.
\end{abstract}

\IEEEpeerreviewmaketitle

\section{Introduction}
\label{introduction}



Technological advances in the last decades have brought about profound changes to the network architecture and operation of Industrial Control Systems (ICS)~\cite{miller2021looking}. ICS is an umbrella term that refers to a collection of interconnected control systems consisting of Supervisory Control and Data Acquisition (SCADA) systems and Distributed Control Systems (DCS). ICS operates industrial processes in critical infrastructures, including energy plants, transportation, and large-scale industrial factories. With the emergence of the Internet of Things (IoT),  Industrial IoT (IIoT), cloud computing, and fifth-generation cellular network (5G) - ICS have become more interconnected, robust, and exposed to the Internet \cite{alladi2020industrial}. From a security point of view, there has been a surge of ransomware and Distributed Denial of Service (DDoS) attacks against ICS in critical infrastructures. For example, the Colonial Pipeline attack in May 2021, causing the largest American pipeline operator to shut down its entire fuel distribution network for six days, sparking significant gasoline shortages~\cite{hobbs2021colonial}.

Since ICS are more interconnected than ever before, attacks not only cause significant economic damage for the affected organisations but are also likely to impact other connected systems and processes in the supply chain. This highlights the need for cooperation and collaboration between ICS operators, governments, and industry groups as a mean to collectively reduce risks and facilitate early prevention of attacks~\cite{bhamare2020cybersecurity}. 
Cyber Threat Intelligence (CTI) sharing is a promising information exchange strategy that is used to collate evidence-based knowledge about cyber threats in terms of potential and current attacks, Indicators of Compromise (IoC), attackers’ motivations, intentions, characteristics and attack vectors, along with their techniques, tactics, and procedures (TTPs)~\cite{jadidi2020securing}. 
While it is widely argued that CTI exchange is a highly valuable tool to secure systems and networks, studies have observed that organisations tend to refrain from engaging in this activity for many reasons e.g., fear of exposure, reputational damage, and lack of incentives for sharing the CTI among one another~\cite{mavroeidis2017cyber}.  


To encourage CTI sharing in ICS, we argue that the use of blockchain is beneficial. Given the blockchain’s salient features of decentralisation, immutability,   security, and auditability, there is a fast-growing desire to use blockchain in developing a network for CTI sharing~\cite{riesco2020cybersecurity}. However, to the best of our knowledge, no attempts have been made that include blockchain for CTI sharing in ICS for fair sharing of high-quality CTI data. 
In this paper, we introduce an approach for CTI sharing in ICS that leverages the advantages of blockchain. Our motivation consists of the following two objectives: i) to examine the need for blockchain-enabled CTI sharing in ICS, and ii) to develop an incentivised framework for the secure dissemination of CTI related to ICS using smart contracts. Our approach motivates better engagement in the exchange of CTI for ICS by employing incentives for entities willing to participate in CTI sharing and their distribution over the blockchain. The major contributions of the paper are: 

\begin{itemize}
    \item We present a comprehensive literature study on existing CTI-sharing solutions and show their critical issues.
    \item We propose a blockchain-enabled CTI sharing framework for ICS offering incentives to participate in the information exchange process. 
    \item We provide a complete design of the system and through various use-case scenarios, demonstrate the suitability of the proposed framework in real-world scenarios. 
\end{itemize}




\begin{table*}[t]
\small
\centering
\caption{Previous proposals on blockchain-enabled CTI sharing and their comparison with our work.}
\label{tab:comparison-previous-works}

\scalebox{.92}{
\begin{tabular}{c*{9}{c}r}
\hline  

Ref. & ICS & Decentralised & Incentives & Legal & Standardised & Quality & Permissioned & Identity & Privacy and \\ {} & Focused & {} & {} & Reporting & Format &  Assurance &  Network &  Verification & Confidentiality \\[0.5ex] 
\hline

\cite{riesco2020cybersecurity} & \xmark & \checkmark & \checkmark & \xmark & \checkmark & \xmark & \xmark & \xmark & \xmark \\ [0.5ex] 

\cite{buber2020blockchain} & \xmark & \checkmark & \xmark & \xmark & \xmark & \checkmark & \xmark & \xmark & \xmark \\ [0.5ex] 

\cite{homan2019new} & \xmark & \checkmark & \xmark & \xmark & \checkmark & \xmark & \checkmark & \checkmark & \checkmark \\ [0.5ex] 

\cite{moubarak2021dissemination} & \xmark & \xmark & \xmark & \xmark & \checkmark & \xmark & \checkmark & \checkmark & \checkmark \\ [0.5ex] 

\cite{preuveneers2020distributed} & \xmark & \xmark & \xmark & \xmark & \xmark & \xmark & \checkmark & \checkmark & \checkmark \\ [0.5ex] 

\cite{he2020blotisrt} & \xmark & \checkmark & \checkmark & \xmark & \checkmark & \checkmark & \checkmark & \checkmark & \xmark \\ [0.5ex] 

\cite{purohit2020defensechain} & \xmark & \checkmark & \checkmark & \xmark & \xmark & \checkmark & \checkmark & \checkmark & \xmark \\ [0.5ex] 

\cite{menges2021dealer} & \xmark & \checkmark & \checkmark & \checkmark & \xmark & \checkmark & \xmark & \checkmark & \xmark \\ [0.5ex] 

\cite{gong2020blocis} & \xmark & \checkmark & \checkmark & \xmark & \xmark & \xmark & \xmark & \xmark & \xmark \\ [0.5ex] 

\cite{gonccalo2020architecture} & \xmark & \checkmark &  \xmark & \xmark & \xmark & \checkmark & \checkmark & \checkmark & \checkmark \\ [0.5ex] 

\textbf{[Our Work]} & \checkmark & \checkmark & \checkmark & \checkmark & \checkmark & \checkmark & \checkmark & \checkmark & \checkmark \\ [0.5ex] 

\hline
\end{tabular}
}
\end{table*}

\section{Related Work}
\label{related-work}
There are several existing solutions for CTI sharing that leverage blockchain. In Table~\ref{tab:comparison-previous-works}, we provide a summary of those proposals in comparison with our current work. 
In~\cite{buber2020blockchain}, the authors present a decentralised threat information sharing system using smart contracts. While it ensured the relevance and usability of information and decentralisation of decision making, it did not provide any design for implementation. Further, unlike our approach, their system has not considered the need for economic incentives to encourage participation, a standardised format for CTI, legal reporting requirements, permissioned network requirement, identity verification, and privacy and confidentiality of shared CTI. The authors in~\cite{homan2019new} suggested a blockchain-based CTI-sharing network using partitioned ‘channels’ based on Traffic Light Protocol (TLP). However, once again, no design consideration was discussed. It also lacked incentives and legal reporting features.


In~\cite{moubarak2021dissemination} and~\cite{preuveneers2020distributed}, two blockchain-based CTI sharing solutions were proposed, each of which recorded transactions and CTI data in a permissioned network, thus achieving identity verification, verifiability and accountability of transactions, and confidentiality. However, unlike our motivation, these solutions failed to achieve decentralisation and depended on a centralised entity that issued decryption keys and certificates. 


Proposals \cite{riesco2020cybersecurity}, \cite{he2020blotisrt}, and \cite{purohit2020defensechain} discuss an incentive-based approach to facilitate trustworthy CTI exchange. However, unlike ours, they did not engage with current discourses on legal reporting obligations, which we incorporate in our paper. They also did not consider the requirement for channels in which critical CTI could be shared privately between two or more specific network members. 


In~\cite{menges2021dealer}, the authors proposed a CTI-sharing application prototype implemented on a public blockchain platform. It facilitated the fulfilment of legal reporting obligations and provided positive token-based incentives for reciprocated engagement in the CTI exchange. It also involved a Critical Infrastructure Component that allowed incident owners to report information to legal authorities. While this component effectively addresses the requirement of legal reporting, the authors suggest that the platform would be used for the exchange of non-critical data only, making it unsuitable for the exchange of highly critical and sensitive CTI data about ICS. Further, unlike ours, their solution is implemented in a permissionless network. Similar to~\cite{menges2021dealer}, proposal~\cite{gong2020blocis} sought to incentivise CTI contribution with tokens and implement their solution in a permissionless blockchain network, which is not appropriate for ICS due to confidentiality and privacy concerns.

In~\cite{gonccalo2020architecture}, a permissioned blockchain network architecture was proposed to facilitate CTI exchange in a trusted environment. CTI contributors were rewarded with increases in reputation level - the higher the reputation level, the more reliable was the CTI shared. Although the study briefly suggested that contributing participants are rewarded with access to `privileged intelligence', it did not explore the different levels of reading access among the different reputation levels. Further, while the architecture used a permissioned blockchain to restrict membership and public and private channels for private sharing, unlike ours, it did not include an identity verification mechanism to prevent infiltration by malicious users.

Overall, existing works have only focused on how blockchain could be used to facilitate CTI exchange within the field of cyber security generally and failed to address the needs and circumstances of ICS operators and relevant stakeholders in the critical infrastructure sector. In comparison, our proposed framework aims to comprehensively address all identified design requirements for sharing CTI about ICS. Specifically, we focus on facilitating the secure exchange of highly sensitive CTI about ICS in a decentralised, permissioned blockchain network. It also enables fulfilment of legal reporting obligations e.g., those imposed upon the critical infrastructure sector in the newly passed Australian \textit{Security Legislation Amendment (Critical Infrastructure) Bill 2020} (Cth), 
as well as leveraging the TLP to allow users to privately collaborate in channels. Moreover, our solution overcomes the financial challenges around the use of tokens in existing solutions by opting for discount-based rewards to encourage active sharing of CTI, thus making CTI more accessible to small and medium-sized enterprises. 

\section{The Proposed Framework}

In this section, we present the proposed framework of CTI sharing in ICS using blockchain and describe its various components and interactions between the components. 

\subsection{System Components}
\begin{figure}[t]
 \centering
    \includegraphics[scale=1.5]{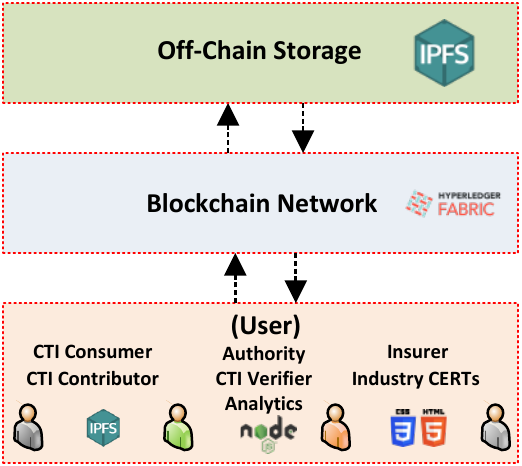}
    \caption{The proposed Cyber Threat Intelligence (CTI) sharing framework.} 
    \label{fig:1}
    \small
\end{figure}

 As shown in Fig.~\ref{fig:1} the proposed framework consists of three components which are:  (i) \textit{off-chain }storage, (ii) \textit{on-chain} blockchain network, and (iii) users. The \textit{off-chain} storage component provides decentralised integrity-secured storage of content-addressable exchange processes, CTI data, and decryption keys through the use of Distributed Hash Tables (DHTs). For \textit{blockchain network}, our design uses a permissioned private blockchain as opposed to a permissionless public blockchain so that users can be uniquely identified and authenticated. The network handles user registration process, CTI contribution and verification process, and CTI consumption process. This component also manages \textit{on-chain} transaction information and executes smart contract chain codes to enable private exchange of CTI among subgroups of users. \textit{Users} are the entities that participate in the network, each of which operates a node and interacts with a user interface to share, verify, and consume CTI. 
The various users’ roles and their associated functions are described as follows: 

\begin{itemize}
\item CTI Consumer: a participating individual or organisation that uses the network to find and consume CTI.
\item CTI Contributor: a participating individual or organisation that uses the network to contribute or share their CTI with other users.
\item Authority: a government authority e.g., the Australian Cyber Security Centre (ACSC) or the Australian Signals Directorate (ASD) that uses the network to receive reports of ICS incidents.
\item Insurer: an insurance company that needs access to CTI to assess a cyber insurance claim following a specific incident.
\item Industry CERTs: industry associations and Cyber Emergency Response Teams (CERTs), which provide advice and support on cyber threats and vulnerabilities to organisations affected by cyber security incidents. 
\item CTI Verifier: a verified independent security expert or researcher who evaluate the quality of CTI shared by a CTI Contributor for the purpose of quality assurance. 
\item Analytics: security analytics providers that can collect and analyse reported CTI to determine attack trends.   
\end{itemize}
Note, a user can hold one or many roles based on the requirements. For example, they can be both a CTI Consumer and a CTI Contributor. Likewise, a CTI Verifier can be a CTI Contributor simultaneously. However, they will not be allowed to verify their own CTI.

\subsection{System Functionality}
As depicted in Fig.~\ref{fig:1}, the implementation of the proposed framework will be made up of three core components, they are (i) the IPFS (InterPlanetary File System) distributed database \cite{ipfs} for \textit{off-chain} storage, (ii) the Hyperledger Fabric network \cite{hyperledger} for the \textit{blockchain} network, and (iii) a decentralised application based on Node.JS, and a user interface powered by IPFS, HTML, and CSS for the \textit{users}. We follow the specific design choice of~\cite{menges2021dealer}. 


The interaction between the IPFS-based database and the Hyperledger Fabric network is bidirectional. Encrypted CTI data is uploaded from the network to the database, while the IPFS returns a unique hash for each piece of CTI data to be stored on the network. Requests for the stored CTI data are sent from the user interface to the blockchain network, which retrieves the data based on their hash from the IPFS database. Depending on user roles, the interactions between the network and the users can be bidirectional or unidirectional. For instance, an insurer might be in the network to consume CTI for the purpose of claim assessment only. A CTI Verifier receives stored CTI for evaluation from the network, and in turn, uploads their evaluation report back to the network.  


In our development, we follow the concept of TLP within the Hyperledger Fabric network as discussed in~\cite{homan2019new}. In particular,~\cite{homan2019new} uses \textit{red}, \textit{amber} and \textit{green} for information disclosure. \textit{Red} indicates access restriction to pre-authorised users only. \textit{Amber} denotes restricted access to the users in the channel only. Finally, \textit{green} allows restricted disclosing of data to all users in the network. We enhance this concept further with a \textit{white}  channel, where data can be freely disclosed to entities outside the network without any restrictions. This is because disclosing incident information and threat warnings about ICS to the public is strongly in the public interest, because ICS are commonly part of critical infrastructures that exist to provide vital services to the population, destruction of which would compromise national security, public health, economy, and public safety. 


\subsection{Technology Selection}
\textbf{Blockchain:} As noted earlier, we plan to use a permissioned private blockchain as opposed to a permissionless public blockchain. Although we follow the design choice of \cite{riesco2020cybersecurity} and \cite{menges2021dealer}, our design uses Hyperledger Fabric instead of the EOS public blockchain or Ethereum for multiple reasons. For example, the proposed solution requires participants to be identified and approved before being given user credentials to access the network. Since the number of users is restricted and transactions are only visible to permissioned users, a permissioned platform like Hyperledger Fabric achieves the goals of confidentiality and privacy. Further, channel support is required to ensure the privacy of transaction data through the implementation of the TLP, which EOS and Ethereum do not support. Moreover, the platform’s consensus mechanism can be flexibly customised to meet the users’ shared goals and agreed governance rules. Furthermore, Hyperledger Fabric does not require the use of a cryptocurrency like EOS or Ethereum, which fits the proposed solution’s incentive model that involves only the registration and subscription fees. Lastly, compared to EOS and Ethereum, Hyperledger Fabric has higher scalability due to having fewer nodes authorised to verify and manage transactions. 

\textbf{Data Storage:} Due to the blockchain's immutability and append-only nature, storing transactions data directly on the blockchain can incur high maintenance costs and prevent new nodes from joining the network \cite{zheng2018innovative}. The proposed network uses  IPFS as the off-chain storage system. As opposed to the traditional location-based addressing method, IPFS uses content-based addressing.  Only the data's unique IPFS transaction hash, which is only 46 bytes long \cite{kumar2019implementation}, is stored in the blockchain, reducing the storage space significantly. With IPFS in the proposed network, one node can request a copy of the stored transaction data from another node, as long as they have access to its unique IPFS hash. The hash is updated every time the transaction content is changed. As long as the hash of the copy matches the hash of the originally stored data, the transaction data's integrity is assured. 




\subsection{Process of CTI Sharing}
\begin{figure}[t]
 \centering
    \includegraphics[width=8.9cm, height=5.5cm]{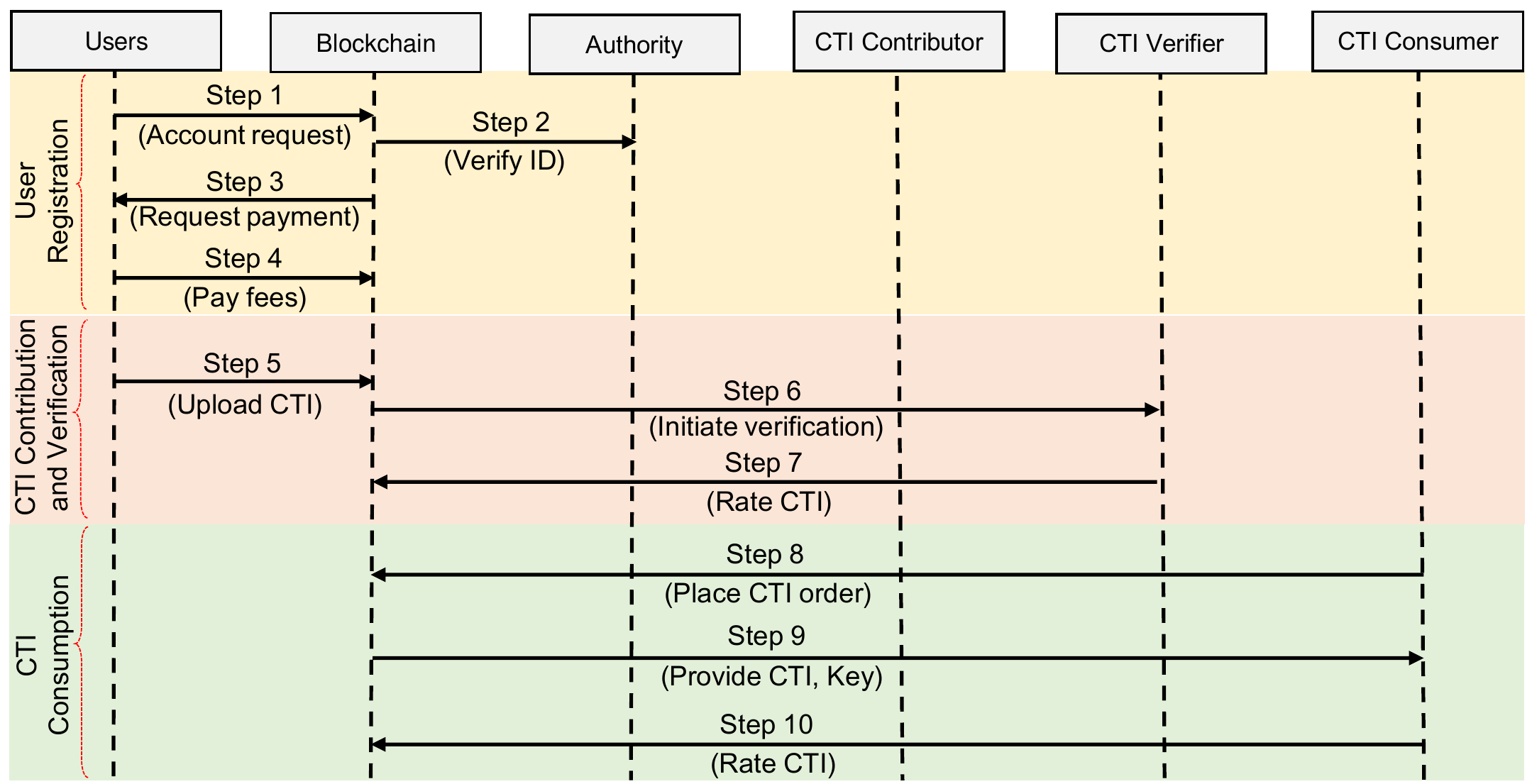}
    \caption{The proposed Cyber Threat Intelligence (CTI) sharing process.}
    \label{fig:flow}
    \small
\end{figure}

In our framework, the CTI sharing process is divided into the following three phases, as illustrated in Fig.~\ref{fig:flow}.

\textbf{User Registration:} 
\begin{itemize}
    \item \textbf{Step 1:} A potential user sends a request to blockchain to create an account. The blockchain requests the user to provide identification (ID) to prove that the user is an ICS operator and/or fits into the description of any of the user roles. The identification can be either a tax file number, identity documents like driver license, business name and number, or official government document. 
    
    \item \textbf{Step 2:} The blockchain sends the provided ID to the Authority for verification. If a user wants to become a Verifier, they must obtain Verifier Access. The blockchain will request the potential Verifier to provide identity documents and proof of their qualifications and expertise to be a Verifier e.g., industry certifications, degrees, 
    and a Verifier Certificate that has been granted by the Authority. 
    
    \item \textbf{Step 3:} If ID verification is successful, the blockchain requests payment for a fixed subscription fee and initial account registration fee from the user. Subscription can be paid on a monthly or annual basis. The fees help to prevent sybil attacks, because they represent a sizeable accumulate amount of financial costs for malicious actors seeking to infiltrate the network. The amount is modest enough to attract small and medium-sized organisations with small budgets. Since every user is required to pay the periodic subscription fees for as long as they want unlimited access to CTI, the `free-rider' problem is also alleviated. 
    
    \item \textbf{Step 4:} The user pays. If payment is successful, the blockchain creates a user account, with the user's chosen username and password. The user may choose to stay anonymous to the other users in the network. Their physical identity is only known to the Authority (with the unique ID provided at Steps 1 and 2). 
\end{itemize}

\textbf{CTI Contribution and Verification:}
\begin{itemize}
    \item \textbf{Step 5:} A CTI Contributor uploads pre-processed CTI to the blockchain. Pre-processing is an internal process at the user’s end that involves formatting data into human-readable format, anonymising personal information, adding descriptive metadata and TLP, and encrypting the CTI. The blockchain stores the CTI with a timestamp.
    
    \item \textbf{Step 6:} The blockchain initiates verification process. Three random Verifiers are selected, each is given a copy of the CTI file. Each copy is encrypted with a different symmetric key that is set up by the CTI Contributor. The CTI Contributor first encrypts the three copies of the CTI with symmetric keys (just one key for each Verifier ($kv_{1}$, $kv_{2}$, $kv_{3}$), then encrypts the symmetric keys with the Verifiers' public keys before uploading the symmetric keys to the blockchain for distribution to the Verifiers. This ensures that the Verifiers' individual access to the CTI is secure from each other and everyone else. 
If someone gains unauthorised access to one of the Contributor's symmetric keys, they cannot access the symmetric keys to decrypt the CTI without one of the Verifier's private keys.
    
    \item \textbf{Step 7:} The Verifiers download and decrypt their respective key to the CTI using their private key, then rate the CTI's quality using platform-provided metrics in terms of accuracy, usability, and relevance to ICS 
    and submit their reports to the blockchain. If the CTI is rated as high-quality CTI, the blockchain gives the Verifiers and the Contributors discount on the next subscription fee. If the data is rated as low-quality or duplicated, only the Verifiers receive a discount. The blockchain then uploads the quality CTI to the public marketplace for consumption and executes smart contracts chain codes as per TLP.
\end{itemize}

\textbf{CTI Consumption:} 
\begin{itemize}
    \item \textbf{Step 8:} A CTI Consumer selects a piece of CTI in the marketplace and places an order. 
    \item \textbf{Step 9:} The blockchain retrieves the encrypted CTI from the database and sends the Consumer the encrypted CTI file and either the Contributor’s symmetric key, or one of the Verifiers’ public keys for decryption.
    \item \textbf{Step 10:} The Consumer uses one of the keys to decrypt the encrypted file. If decryption is successful, the CTI Consumer confirms so with the blockchain and gives a rating on the quality of the CTI. If decryption fails, the Consumer requests another key from the blockchain, which then obtains it from the remaining three keys from the Verifiers. 
\end{itemize}

\section{Use Case Scenarios and Discussion}


In this section, we discuss various practical use-case scenarios to examine the suitability of our proposed framework in real-world applications. 

\textit{\textbf{Scenario 1:} A user wants to exchange CTI about ICS in a permissioned blockchain network that provides identity verification and confidentiality, security, and privacy.} 

As discussed above, threat intelligence information about ICS, especially ICS in critical infrastructures like power plants and transportation systems, carries significant values and can have enormous impact on the economy, national security, and public wellbeing. The proposed network is to be implemented in a permissioned decentralised network based on Hyperledger Fabric. Users who do not fall into one of the roles specified and/or cannot provide unique identification and/or insist on being anonymous will not be permitted to join the network. While this could represent a limitation in terms of the CTI that could have been shared with the network by a non-user, the security and privacy of the network is paramount.


Our solution ensures privacy and confidentiality of CTI by requiring that the CTI is encrypted using cryptography, while ensuring that the CTI is shared privately in TLP-supported ‘channels’ with only the intended recipients. While a compromise of the encryption technology might result in compromise of the confidentiality of the data stored on IPFS, this scenario is considered unlikely on the assumption that the encryption technology used will be the latest in the market. This risk is not entirely impossible, since the user organisation itself may be compromised, leading to the compromise of their network credentials. 

\textit{\textbf{Scenario 2:} A user wants to share sensitive CTI data with the Authority as part of their legal reporting obligation in a private channel.}

ICS operators and other organisations that are required to provide CTI to the Authority following an incident can do so in the network by implementing the \textit{red} protocol. The network’s smart contract enforces a check on the TLP applicable to the CTI before it is shared with the Authority. Users that are not the original Contributor or the Authority cannot view or join the private channel. Evidence of compliance with legal reporting can be shown and traced through the timestamp that is included in the CTI metadata. 

\textit{\textbf{Scenario 3:} A user wants access to CTI that follows a standardised format.}

The format of shared CTI in the network is standardised to ensure consistency (by the blockchain's data governance rules, as stated in pre-processing requirement before sharing). 
CTI Contributors are expected to pre-process their data in the required format, with the required metadata, and encrypted before uploading to the network. The standardised format also allows for a future implementation of filtering in the user interface application, which would allow CTI Consumers to browse and search the marketplace for the most relevant CTI. For example, they will be able to narrow the search results by filtering by industry, type of ICS, vulnerability, attacks, etc.  

\textit{\textbf{Scenario 4:} A user wants to share CTI only with users that have been uniquely identified and authorised to participate in the network, not Sybil malicious accounts.}

The framework requires all users to prove their physical identity with unique identity documents (e.g., proof of business name and ownership, or official government documents) 
and pay a registration fee followed by periodic subscription fees before joining the network. These rigorous measures present significant barriers for malicious actors to create Sybil accounts, thus satisfying the permission requirement and securing the network against Sybil attacks. 

\textit{\textbf{Scenario 5:} A user wants to have access to CTI at an affordable price and be rewarded economically for sharing CTI, thus encouraging reciprocated exchanges among users. }

Compared to existing CTI-sharing platforms that incentivise voluntary sharing through the use of tokens and transaction fees, our solution is more inclusive of stakeholder organisations that may not have the financial capability to pay for every piece of CTI. Alternatively, organisations may be reluctant to use tokens, given the volatility and uncertainty of the cryptocurrency market. Our solution opts for a sufficient but modest amount of periodic subscription fees instead, which provides organisations unlimited access to CTI. This way, the cost for CTI is made fairer, no matter the size of the participating organisation. In terms of economic incentives, the platform offers discounts on the periodic subscription fees in return for quality CTI, which can encourage users to share more CTI and enable them to offset the costs incurred in the threat detection, data collection, and incident recovery processes. Verifiers also receive the discounts for ensuring the quality of CTI.

\textit{\textbf{Scenario 6:} A user wants access to high-quality CTI.}
Our solution provides quality assurance in the verification process. Each piece of CTI must be evaluated by three independent expert Verifiers before they are uploaded to the marketplace. The Verifiers are selected at random, which can make Verifiers’ collusion challenging. Verifiers’ free-riding is prevented because CTI Consumers also provide a CTI rating, which will be cross-checked with Verifiers’ reports should there be a discrepancy. Verifiers are likely to be reluctant to collude or submit fraudulent ratings, because they are physically identified and can be removed from the network through majority voting. The technical expertise of Verifiers is guaranteed through the certification scheme administered by the Authority. 


In summary, the work presented in this paper builds on existing work to explore how a CTI-sharing solution can be designed to motivate better engagement in the exchange of CTI for ICS. 
While existing solutions focused on CTI-sharing in cyber security generally, 
this work focuses on exploring the requirements of stakeholders of ICS specifically. It presents a solution that features several improvements that are tailored to the identified requirements e.g., a permissioned network and private communication channels based on TLP to facilitate legal reporting of incidents and cooperation with government authorities. Moreover, our solution addresses the financial challenges that some small and medium-sized organisations may have by only requiring the payments of registration and subscription fees without any transaction fees, thus making the costs of participation fairer. It also removes uncertainty around the fluctuating value of cryptocurrency tokens and offers subscription fee discount as the more stable, trustworthy alternative of economic incentive for CTI contribution.

\section{Conclusion and Future Work}

This paper explores the factors influencing organisations’ decisions to exchange CTI about ICS and proposes a network model that addresses such factors to encourage the secure dissemination of CTI related to ICS. There is not yet a platform that enables ICS stakeholders to engage in CTI exchange efficiently and effectively. We argued that blockchain is a promising solution for the development of such an ICS-focused platform. This study, therefore, examined the designs of existing solutions to identify features that would suit the needs of ICS stakeholders. We proposed a framework based on Hyperledger Fabric blockchain and IPFS distributed storage built on and extended previous work. The proposed framework can facilitate organisations’ compliance with legal reporting obligations and promote the collaborative exchange of CTI among ICS stakeholders using discount-based incentives and quality assurance mechanisms through expert verification. The network was designed with confidentiality and privacy in mind through the application of Traffic Light Protocol using Hyperledger Fabric’s channel capabilities, which allowed users to exchange CTI in private sub-groups securely. The framework was also designed with flexibility in mind, in the sense that a permissioned blockchain network that provides similar features to those offered by Fabric can also be used to implement it. However, Fabric has been shown to be one of the most innovative enterprise-grade blockchain solutions in the market. We have examined the proposed framework's suitability, relevance, and feasibility in real-world applications scenarios. 
In future work, the solution will still be implemented on Hyperledger Fabric to evaluate its performance, viability, and practicality. Future work will also investigate which quality evaluation metrics the CTI Verifiers must use. 

\ifCLASSOPTIONcaptionsoff
  \newpage
\fi

\bibliographystyle{IEEEtran}
\bibliography{bare_jrnl}

\end{document}